\documentclass[journal]{IEEEtran}
\usepackage{color}

\usepackage{cite}
\usepackage{graphicx}
\graphicspath{{./figs/}}
\DeclareGraphicsExtensions{.pdf,.jpeg,.png}

\usepackage{amsmath}
\begin{document}
\title{Robustness Investigation on Deep Learning CT Reconstruction for Real-Time Dose Optimization}

\author{Chang Liu, Yixing Huang, Joscha Maier, Laura Klein, Marc Kachelrieß, Andreas Maier
%\thanks{Manuscript received May X, 2020}% <-this % stops a space
\thanks{C. Liu, Y. Huang and A. Maier are with Pattern Recognition Lab, Friedrich-Alexander-University Erlangen-Nuremberg, Erlangen, Germany (email:
chang.ch.liu@fau.de).}
\thanks{J. Maier, L. Klein and M. Kachelrieß are with German Cancer Research Center, Heidelberg, Germany}
%\thanks{A. Maier is also with Erlangen Graduate School in Advanced Optical Technologies (SAOT), Erlangen, Germany.}% <-this % stops a space
%\thanks{Corresponding author is Chang Liu, Email: chang.ch.liu@fau.de}}
}
\maketitle

\begin{abstract}
In computed tomography (CT), automatic exposure control (AEC) is frequently used to reduce radiation dose exposure to patients. For organ-specific AEC, a preliminary CT reconstruction is necessary to estimate organ shapes for dose optimization, where only a few projections are allowed for real-time reconstruction. In this work, we investigate the performance of automated transform by manifold approximation (AUTOMAP) in such applications. 
%For a proof of concept, several models were trained using MNIST dataset with different training and testing data organization. The reconstructions are then evaluated in both qualitative and statistical way. In the end the AUTOMAP sparse-view reconstruction is tested using chest CT tomograms. We analyse the pros and cons of AUTOMAP and hopefully provide the mitigation method.
For proof of concept, we investigate its performance on the MNIST dataset first, where the dataset containing all the 10 digits are randomly split into a training set and a test set. We train the AUTOMAP model for image reconstruction from 2 projections or 4 projections directly. The test results demonstrate that AUTOMAP is able to reconstruct most digits well with a false rate of 1.6\% and 6.8\% respectively. In our subsequent experiment, the MNIST dataset is split in a way that the training set contains 9 digits only while the test set contains the excluded digit only, for instance “2”. In the test results, the digit ``2"s are falsely predicted as ``3" or ``5" when using 2 projections for reconstruction, reaching a false rate of 94.4\%. For the application in medical images, AUTOMAP is also trained on patients' CT images. The test images reach an average root-mean-square error of 290 HU. Although the coarse body outlines are well reconstructed, some organs are misshaped.
\end{abstract}


\section{Introduction}

%\IEEEPARstart{T}{his} demo file is intended to serve as a ``starter file''

\IEEEPARstart{C}{omputed} tomography (CT) is widely used for disease diagnosis and interventions in modern medicine. 
%The major risk of CT scans to patients is the relatively high radiation dose. The dose of one CT scan is estimated to be 10 to 100 times higher than one radiography \cite{suetens_2009}, thus the dose is not negligible and is expected to be reduced. 
To reduce health risks caused by X-rays, automatic exposure control (AEC) is frequently used in contemporary CT scanners to control the effective dose as low as reasonably achievable while keeping the image quality. 
%Conventional AEC methods adapt the tube current based on the patient size, angular and z-axis position \cite{kubo2008radiation}. 
For accurate organ-specific AEC, a preliminary reconstruction of patient organs is necessary for dose optimization, which needs to be fast enough for real-time AEC. Therefore, a reconstruction from extreme sparse-view projections, e.g. 4 projections or 2 projections, is preferred.

%the applied dose should be optimized based on the dose received by organs. A coarse estimation of the CT volume is then necessary before the scan and a fast sparse-view reconstruction method is preferred to carry out such patient-specific AEC.  

For image reconstruction from sparse-view data, conventional filtered back-projection (FBP) based reconstruction algorithms perform poorly. Even for compressed sensing technologies, no satisfactory images are reconstructed from such extreme sparse views. However, the emerging deep learning methods offer a possible solution. Some achievements have been reported to reconstruct images directly from projection data by deep learning \cite{wurfl2016deep,zhu2018image,li2019learning,ying2019x2ct}. Automated transform by manifold approximation (AUTOMAP) \cite{zhu2018image} is a generic deep learning reconstruction framework for multiple imaging modalities including CT. It learns a joint manifold linking projections and reconstructed images in a supervised manner. As the robustness of deep learning is a concern for clinical applications \cite{huang2018some}, in this paper, we investigate the performance of AUTOMAP for such extreme sparse-view reconstruction.

\section{Methods}
According to \cite{zhu2018image}, the AUTOMAP neural network is implemented for parallel-beam CT reconstruction from 4 projections or 2 two projections. As a proof of concept, the performance of AUTOMAP is evaluated on the MNIST dataset first. Afterwards, its performance is evaluated on patients' CT images.

\subsection{Neural network architecture}
The AUTOMAP used in our experiments consists of 3 cascaded fully-connected layers, each with a tanh activation function, and 3 convolutional layers, each with a ReLu activation function. Batch normalization is included to smooth the training process. 

\subsection{Training and test}
For the experiments on the MNIST dataset, 48000 images from MNIST are used for training and test. In our first investigation, we randomly split the 48000 images containing all the 10 digits into 43000 images for training and 5000 for test. The images are first resized to $64\times64$ and encoded using the parallel-beam Radon transform into sinograms with a detector size of 64. The detector pixel size is the same as the image pixel size, simply assuming 5\,mm per pixel. The projections from $0^{\circ}$, $45^{\circ}$, $90^{\circ}$ and $135^{\circ}$ are used as the first sparse-view condition, and those from $0^{\circ}$ and $90^{\circ}$ are used as the second sparse-view condition. In the second investigation, the dataset is split based on the digits. We exclude the images of digit ``2" from the dataset and train the model using remaining images using the same sparse-view conditions. The images of digit ``2" are used for test afterwards. 
%We also repeat this investigation by excluding digit ``8".

To demonstrate the performance of AUTOMAP on medical images, we further train the model using 6900 sinograms from 12 patients' CT volumes. Sinograms from 11 CT volumes are used for training, while those from the excluded one are used for test. The other experimental settings are the same as the first experiment.

For each training, 50 epochs are applied. The loss function used is mean square error (MSE) and the optimizer used is RMSProp with a learning rate of 0.00002.

\subsection{Evaluation}
For the experiment on the MINIST images, the image quality of reconstructed images is evaluated by root-mean-square error (RMSE) and false digit rate. A false digit means that the reconstructed image is observed as a different digit rather than the digit ``2", or it cannot be recognized as a digit at all. For the experiment on CT images, RMSE in Hounsfield unit (HU) is calculated.

\section{Results And Discussion}

\begin{figure}[tbh]
\centering
\includegraphics[width=0.45\textwidth]{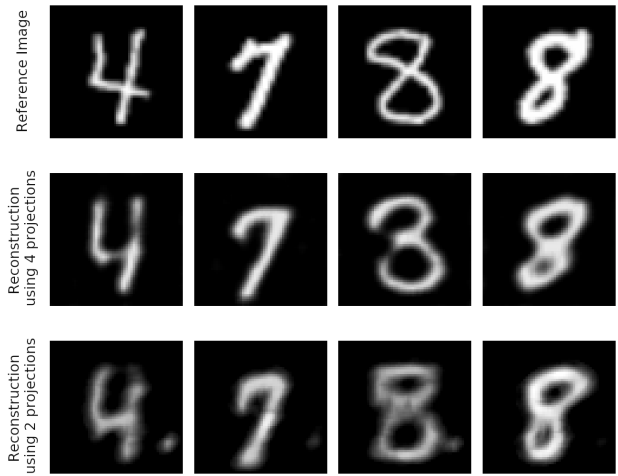}
\caption{Example results of sparse-view reconstruction using AUTOMAP trained on the randomly split MNIST dataset.}
\label{fig_random}
\end{figure}

\begin{figure}[tbh]
\centering
\includegraphics[width=0.45\textwidth]{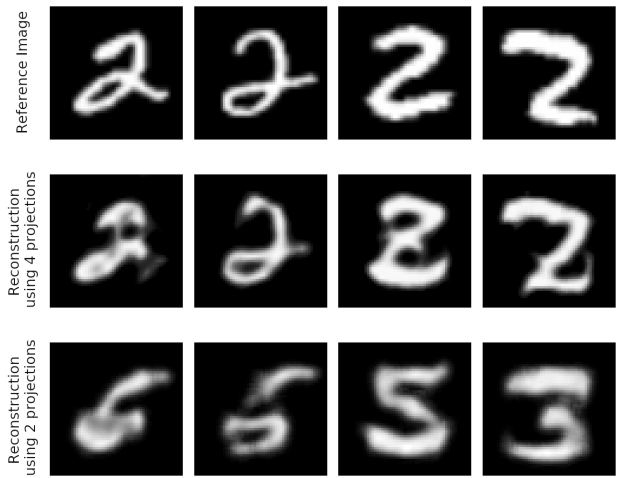}
\caption{Example results of sparse-view reconstruction using AUTOMAP trained on the MNIST dataset excluding digit ``2". The reconstructions from 4 projections are relatively stable. In contrast, those from 2 projections are mostly false digits.}
\label{fig_exclude_2}
\end{figure}

\begin{figure}[tbh]
\centering
\includegraphics[width=0.45\textwidth]{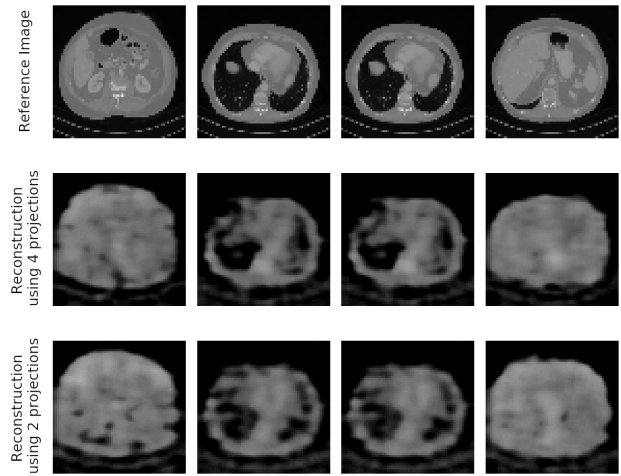}
\caption{Example of sparse-view reconstruction using AUTOMAP trained on CT images.}
\label{fig_ct}
\end{figure}

\begin{table}[tbh]
\centering
\caption{Results of sparse-view reconstruction using MNIST data.}
\label{res_table}
\begin{tabular}{lcccc}
\hline
                                                                & \begin{tabular}[c]{@{}c@{}}4 projs \\ RMSE\end{tabular} & \begin{tabular}[c]{@{}c@{}}2 projs \\ RMSE\end{tabular} & \begin{tabular}[c]{@{}c@{}}4 projs false \\ digit rate\end{tabular} & \begin{tabular}[c]{@{}c@{}}2 projs false \\ digit rate\end{tabular} \\ \hline
\begin{tabular}[c]{@{}l@{}}MNIST \\ randomly split\end{tabular} & 0.123                                                   & 0.169                                                   & 1.6\%                                                               & 6.8\%                                                               \\ \hline
\begin{tabular}[c]{@{}l@{}}MNIST \\ excluding ``2"\end{tabular}  & 0.134                                                   & 0.244                                                   & 8.4\%                                                               & 94.4\%                                                  \\ \hline
\end{tabular}
\end{table}

Fig.~\ref{fig_random} exhibits that AUTOMAP is able to reconstruct most digits well when the training dataset contains all the 10 digits. In Fig.~\ref{fig_exclude_2} where the digit ``2" is excluded for training, most digits are recognized as ``2" well for reconstructions from 4 projections, although one result appears like the letter ``$\mathcal{Z}$". However, most reconstructions appear as false digits, e.g., ``3", ``5" or ``6", when only 2 projections are used.
As shown in TAB.~\ref{res_table}, the false digit rate is 94.4\% when training on the MNIST dataset excluding ``2" using 2 projections.
 The reconstructions on the CT images are displayed in Fig. \ref{fig_ct}. The coarse body outlines are reconstructed by AUTOMAP. However, some organs are misshaped. These reconstructions have an average RMSE value of 290 HU. Whether such images are sufficient for dose optimization in clinical applications needs further investigation.

\section{Conclusion}

As a proof of concept, AUTOMAP shows its robustness for sparse-view reconstruction when using 4 projections on the MINIST dataset. However, it tends to reconstruct false digits from 2 projections when one digit is excluded for training. The experiment on the patients' CT images demonstrates that AUTOMAP is able to reconstruct the coarse body outlines well. However, some organs are misshaped.


\ifCLASSOPTIONcaptionsoff
  \newpage
\fi




\bibliographystyle{IEEEtran}
\bibliography{refs}

\begin{thebibliography}{1}
\providecommand{\url}[1]{#1}
\csname url@samestyle\endcsname
\providecommand{\newblock}{\relax}
\providecommand{\bibinfo}[2]{#2}
\providecommand{\BIBentrySTDinterwordspacing}{\spaceskip=0pt\relax}
\providecommand{\BIBentryALTinterwordstretchfactor}{4}
\providecommand{\BIBentryALTinterwordspacing}{\spaceskip=\fontdimen2\font plus
\BIBentryALTinterwordstretchfactor\fontdimen3\font minus
  \fontdimen4\font\relax}
\providecommand{\BIBforeignlanguage}[2]{{%
\expandafter\ifx\csname l@#1\endcsname\relax
\typeout{** WARNING: IEEEtran.bst: No hyphenation pattern has been}%
\typeout{** loaded for the language `#1'. Using the pattern for}%
\typeout{** the default language instead.}%
\else
\language=\csname l@#1\endcsname
\fi
#2}}
\providecommand{\BIBdecl}{\relax}
\BIBdecl

\bibitem{wurfl2016deep}
T.~W{\"u}rfl, F.~C. Ghesu, V.~Christlein, and A.~Maier, ``Deep learning
  computed tomography,'' in \emph{Proc. MICCAI}, 2016, pp. 432--440.

\bibitem{zhu2018image}
B.~Zhu, J.~Z. Liu, S.~F. Cauley, B.~R. Rosen, and M.~S. Rosen, ``Image
  reconstruction by domain-transform manifold learning,'' \emph{Nature}, vol.
  555, no. 7697, pp. 487--492, 2018.

\bibitem{li2019learning}
Y.~Li, K.~Li, C.~Zhang, J.~Montoya, and G.-H. Chen, ``Learning to reconstruct
  computed tomography images directly from sinogram data under a variety of
  data acquisition conditions,'' \emph{IEEE Trans. Med. Imaging}, vol.~38,
  no.~10, pp. 2469--2481, 2019.

\bibitem{ying2019x2ct}
X.~Ying, H.~Guo, K.~Ma, J.~Wu, Z.~Weng, and Y.~Zheng, ``{X2CT-GAN}:
  reconstructing ct from biplanar x-rays with generative adversarial
  networks,'' in \emph{Proc. IEEE CVPR}, 2019, pp. 10\,619--10\,628.

\bibitem{huang2018some}
Y.~Huang, T.~W{\"u}rfl, K.~Breininger, L.~Liu, G.~Lauritsch, and A.~Maier,
  ``Some investigations on robustness of deep learning in limited angle
  tomography,'' in \emph{Proc. MICCAI}, 2018, pp. 145--153.

\end{thebibliography}

\end{document}